\documentclass{jpsj3}
\usepackage{txfonts}
\usepackage{graphicx}
\usepackage{dcolumn}
\usepackage{color}
\usepackage{array,multirow,makecell}
\usepackage{adjustbox}
\usepackage{bm}
\usepackage{nicefrac}
\usepackage{multicol,blindtext}
\title{Feedback of superconductivity on the magnetic excitation spectrum of UTe$_{2}$}

\author{St\'ephane Raymond$^1$, William Knafo$^2$, Georg Knebel$^3$, Koji Kaneko$^{4,5}$, Jean-Pascal Brison$^3$, Jacques Flouquet$^3$, Dai Aoki$^{6}$, G\'erard Lapertot$^3$}
\inst{$^1$Univ. Grenoble Alpes, CEA, IRIG, MEM, MDN, 38000 Grenoble, France\\
$^2$ LNCMI-EMFL, CNRS UPR3228, Univ. Grenoble Alpes, Univ. Toulouse 3, INSA-T, Toulouse and Grenoble, France \\
$^3$Univ. Grenoble Alpes, CEA, Grenoble INP, IRIG, PHELIQS, 38000, Grenoble, France \\
$^4$Materials Sciences Research Center, Japan Atomic Energy Agency, Tokai, Ibaraki 319-1195, Japan\\
$^5$Advanced Science Research Center, Japan Atomic Energy Agency, Tokai, Ibaraki 319-1195, Japan\\
$^6$Institute for Materials Research, Tohoku University, Ibaraki 311-1313, Japan}

\abst{We investigate the spin dynamics in the superconducting phase of UTe$_{2}$ by triple-axis inelastic neutron scattering on a single crystal sample. At the wave-vector $\bf{k_1}$=(0, 0.57, 0), where the normal state antiferromagnetic correlations are peaked, a modification of the excitation spectrum is evidenced, on crossing the superconducting transition, with a reduction of the relaxation rate together with the development of an inelastic peak at $\Omega$ $\approx$ 1 meV. The low dimensional nature and the the $a$-axis polarization of the fluctuations, that characterise the normal state, are essentially maintained below $T_{sc}$. The high ratio $\Omega/k_{B}T_{sc}$ $\approx$ 7.2 contrasts with the most common behaviour in heavy fermion superconductors.}

\begin{document}
\maketitle

The discovery of the heavy fermion (HF) superconductor compound UTe$_{2}$ ($T_{sc}$ $\approx$ 1.6 K) \cite{Ran1,Aoki1} has triggered a wealth of research owing to the possible triplet and chiral nature of the superconductivity \cite{Jiao} and the observation of multiples superconducting phases under magnetic field and pressure as well as the proximity to a magnetic quantum instability \cite{Braithwaite,Knebel1,Ran2,Aoki2,Ran3,Lin,Knebel2,Thomas,KnafoCom}. In unconventional superconductors, the magnetic excitation spectrum measured by inelastic neutron scattering (INS) is often modified when entering the superconducting state through the occurrence of a new magnetic excitation usually named resonance and common to cuprates, Fe-based superconductors and HF systems \cite{Scalapino}.
The similarity of the resonance in these different systems lies in the two prevalent features : the resonance is observed at specific wave-vectors associated with features of the Fermi surface (i.e. nesting) and/or features of the superconducting gap (i.e. change of sign) and for an energy $\Omega$ related to the superconducting gap $\Delta$ \cite{Yu}.
In HF systems, the intrinsic low energy scales and the more exotic superconducting states inferred form strong spin-orbit coupling associated with $f$-electrons make the observation of the resonance scarce. Indeed, among the numerous HF superconductors \cite{Pfeiderer}, the resonance peak is so far reported for a few compounds only: CeCoIn$_{5}$ \cite{Stock,Raymond}, CeCu$_{2}$Si$_{2}$ \cite{Stockert}, UPd$_{2}$Al$_{3}$\cite{Metoki,Bernhoeft} and UBe$_{13}$ \cite{Hiess}.

Up to now, INS measurements performed in the normal state of UTe$_{2}$ detect only incommensurate spin fluctuations peaked at the wave-vector $\bf{k_1}$=(0, 0.57, 0) \cite{Duan1}.
A typical quasielastic response with a relaxation rate of 2.5 meV and a low dimensional behaviour of the fluctuations together with a strong polarization of the fluctuations along the $a$-axis were further evidenced \cite{Knafo}. Finally, a spin resonance mode was found at an energy of about 1 meV  in the superconducting phase for the wave-vector $\bf{k_1}$ \cite{Duan2}. In the present work, we confirm on a unique single crystal sample, this modification of the magnetic excitation spectrum, from quasielastic to inelastic, on crossing $T_{sc}$ and show that the low dimensional character of the fluctuations along the $c$-axis and their polarization along the $a$-axis are essentially unchanged in the superconducting phase. The different models of resonance for $f$-electron systems are briefly reviewed together with their relevance to our experimental findings in comparison with other HF compounds.

The INS measurements were carried out on the cold neutron three axis spectrometer IN12 \cite{Schmalzl} located at the high flux reactor of the Institut Laue Langevin, Grenoble.
The instrument was operated without its velocity selector and a Be filter was placed in the incident neutron beam. The data were taken with fixed final neutron wavevector $k_F$ = 1.2 ${\AA^{-1}}$ up to an energy transfer of 1.9 meV (corresponding to an initial neutron wavevector $k_I$=1.535 ${\AA^{-1}}$) using the double focusing pyrolitic graphite monochromator and the horizontal focusing pyrolitic graphite analyzer and without collimations. In this paper, the bare neutron intensity is presented normalized to an incident beam monitor count corresponding to an average measurement time of 25 min.
The sample is the same as in Ref.~[\citen{Knafo}] and is a sole single crystal of total mass 241 mg to be compared with the assembly of twenty seven pieces of single crystals (900 mg) used in Ref.~[\citen{Duan2}].
The sample was installed in a helium-3 fridge. The base temperature increased from 0.4 to 0.6 K, on turning the neutron beam on, due to the heating of a Cadmium foil (neutron absorber) placed above the sample.

\begin{figure}
\includegraphics[width=8cm]{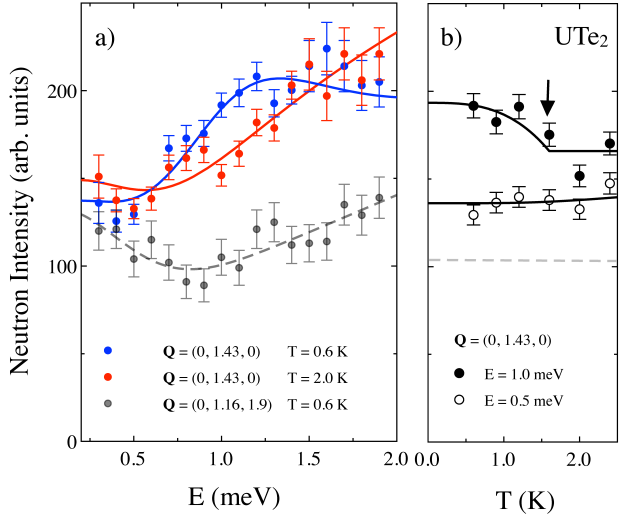}
\caption{a) Energy spectra at $\bf{Q}$=(0, 1.43, 0) at $T$= 0.6 and 2 K and at $\bf{Q}$=(0, 1.16, 0.9). Full (signal) and dash (background) lines are fits as explained in the text. b) Temperature variation of the neutron intensity measured at $\bf{Q}$=(0, 1.43, 0) for energy transfers of 0.5 and 1 meV.  The full lines are fits as explained in the text. The dash line is the background determined from panel a) and similar for 0.5 and 1 meV.}
\label{f1}
\end{figure}

\begin{figure}
\includegraphics[width=7cm,angle=-90]{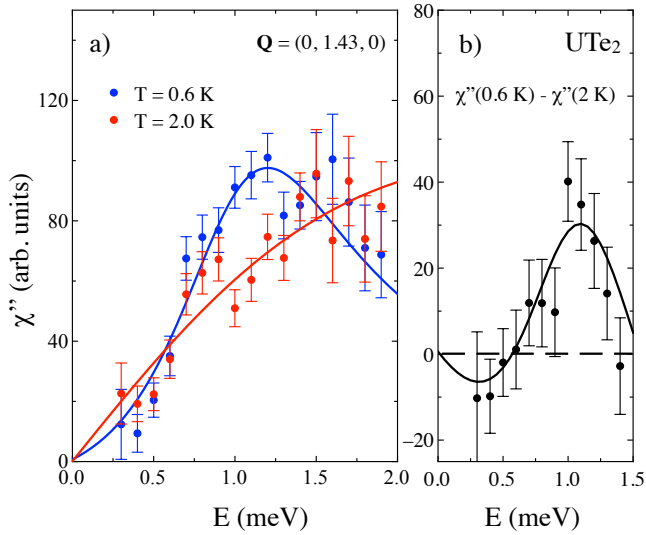}
\caption{a) Imaginary part of the dynamical spin susceptibility at $\bf{Q}$=(0, 1.43, 0) obtained from the data shown in Fig.~\ref{f1}. The lines are the fits described in the text and the same as in  Fig.~\ref{f1}. b) Difference of the data points and of the fits shown in panel a).}
\label{f2}
\end{figure}

\begin{figure}
\centering
\includegraphics[width=7cm]{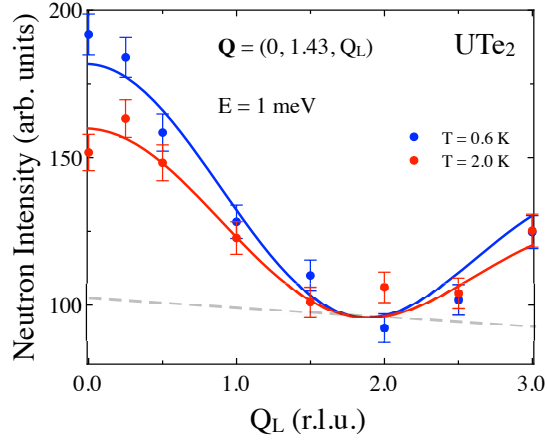}
\caption{Constant energy scan performed at 1 meV along $\bf{Q}$=(0, 1.43, $Q_L$) at 0.6 and 2 K. The solid lines are fits as explained in the text. The dash line indicates the background.}
\label{f3}
\end{figure}

The magnetic excitation spectrum measured at the scattering vector $\bf{Q}$=(0, 1.43, 0) corresponding to the wavevector $\bf{k_1}$ ($\bf{Q}$=(0, 2, 0)-$\bf{k_1}$) is shown in Fig.~\ref{f1}a) at 0.6 and 2 K.
The background measured at 0.6 K and $\bf{Q}$=(0, 1.16, 0.9), a scattering-vector far from any magnetic correlations\cite{Knafo} and having the same modulus as $\bf{Q}$=(0, 1.43, 0), is also shown.
This background is phenomenologically fitted by the sum of a Gaussian peak centred at zero energy (incoherent signal tail) and a sloping background.
At 2 K, the magnetic signal is conveniently described by a quasielastic Lorentzian lineshape taking into account the temperature population factor and using the background determined above.
The obtained relaxation rate in this normal state, $\Gamma_{n}$= 3.0 (7) meV, is consistent with our previous report of measurements performed with a worse resolution in a higher energy range (0.6 to 7.5 meV) \cite{Knafo}.
In contrast, at 0.6 K the magnetic excitation spectrum cannot be described by this simple relaxational response due to the transfer of spectral weight from low energy (below 0.5 meV) to high energy (0.75-1.5 meV) region.
In order to keep a consistent description, the spectrum below $T_{sc}$ is conveniently described by adding a pole at an energy $\Omega$ in the above quasielastic response and by taking into account Stokes (pole at +$\Omega$) and anti-Stokes (pole at -$\Omega$) peaks (See e.g. Ref.~[\citen{Panarin}]).
The obtained inelasticity is $\Omega$=1.00 (4) meV and the damping in the superconducting state is $\Gamma_{sc}$=0.68 (7) meV.
Figure \ref{f1}b shows the temperature dependence of the neutron intensity measured at $\bf{Q}$=(0, 1.43, 0) for constant energy transfers of 0.5 and 1 meV.
At 0.5 meV, the intensity slightly decreases when decreasing temperature and is described at first approximation by the temperature population factor (solid line in Fig.\ref{f1}b).
In contrast, the intensity at 1 meV increases markedly below $T_{sc}$ and is phenomenologically described by an order parameter-like variation with a fixed transition temperature at 1.6 K.
In order to better highlight the evolution of the magnetic excitation spectrum on crossing $T_{sc}$ and to distinguish intrinsic behaviour from thermal effects, the imaginary part of the dynamical spin susceptibilty, $\chi"(\mathbf{k_1},E)$, obtained by subtracting the fitted background and taking into account the temperature factor is plotted in Fig.~\ref{f2}a. At high temperature, the linear increase of $\chi"$ is characteristic of the quasielastic Lorentzian response for energies much smaller than the relaxation rate (slope $\chi'$/$\Gamma_n$).
In contrast this typical response is not seen in the low temperature dynamical susceptibility where the data below 0.6 meV are systematically lower than the one in the normal phase and the one in the range 0.6-1.4 meV are significantly higher. This redistribution of spectral weight is also highlighted in the the difference of $\chi"(\mathbf{k_1},E)$ taken between 0.6 and 2.0 K and shown in Fig.~\ref{f2}b. Altogether the data point to the suppression of the low energy excitations and the development of a well-defined mode in the excitation spectrum.
Figure \ref{f3} shows a constant energy scan performed along $\bf{Q}$=(0, 1.43, $Q_L$) for an energy transfer of 1 meV at 0.6 and 2 K.
The line is a fit to the $Af_m^2(\mathbf{Q})cos^2(\pi Q_Ld_1/c)$ modulation introduced in Ref.~[\citen{Knafo}], where $f_m$ is the uranium magnetic form factor. This wave-vector dependence reflects the in-phase fluctuations of the two uranium atoms of the unit cell separated by distance $d_1$ along the $c$-axis together with i) the absence of correlations along the $c$-direction and ii) the dominant polarization along the $a$-axis of the fluctuations. In the ladder structure of UTe$_{2}$, $d_1$ corresponds to the rung length. Except for a difference in the amplitude $A$ [$A$(0.6 K)=129(6) arb. units and $A$(2 K)=94(5) arb. units], all other parameters being kept fixed, the same modulation is found to describe the data at 0.6 and 2 K (a sloping background is used to describe all the data consistently). For completeness, the absence of elastic scattering corresponding to a static ordering was checked at $\bf{Q}$=(0, 1.43, 0) for 0.6 and 2 K (data not shown here).

Our data on the excitation spectrum at $\bf{k_1}$ in the superconducting state of UTe$_{2}$ are consistent with the report of Ref.~[\citen{Duan2}].
The broad nature of the excitation mode is confirmed using the three-axis spectrometer technique on a unique single crystal for which the half width at half maximum of the (0, 0, 4) Bragg peak rocking curve profile is 1.7 degree on IN12.
In contrast, the data of Duan et al., were obtained on an assembly of single crystal samples and with a significant averaging in wave-vector and energy space inherent to the use of the time of flight spectrometer technique\cite{Duan2}. Regarding the wave-vector dependence of the excitation, Duan et al. focused on the $a$ and $b$ directions while we focused on the $c$-axis. 
Both studies show that the wave-vector dependence measured for the maximum intensity signal at 1 meV is similar between the normal and the superconducting phase, with an enhanced amplitude in the latter phase. 
The resonance in UTe$_{2}$ is canonically located at the wave-vector where the normal state paramagnetic correlations built up. The unusual feature in comparison with other HF systems is its rather high energy compared to $k_{B}T_{sc}$ ($\Omega/k_{B}T_{sc}$ $\approx$ 7.2) and also the important damping in the superconducting state $\Gamma_{sc}$/$\Omega$ $\approx$ 0.7. 

In INS experiments performed on unconventional superconductors, two features characterize the excitation spectrum.
The first feature is the a suppression of spectral weight for energies below 2$\Delta$, where $\Delta$ is the superconducting gap. The effect has its counterpart in conventional superconductors, see e.g. the suppression of phonon damping below 2$\Delta$ \cite{Shapiro} or the suppression of the relaxation of crystal field levels \cite{Feile}.
The second effect specific to unconventional superconductors is the appearance of a sharp resonance mode at low energy most often below 2$\Delta$.  
Several models are discussing the resonance phenomena, based either on an itinerant or localized description of the magnetic excitation spectrum.
In the most common model, the so-called spin-1 exciton \cite{Eschrig}, the Fermi surface topology is playing an essential role and the normal state susceptibilty can already be boosted by nesting effects.
For a superconducting gap with a change of sign in the vicinity of such vector, the susceptibility is further enhanced by a BCS-like coherence factor.
Finally, when residual interactions are taken into account, a collective spin-1 exciton mode pushed below the continuum of electron-hole excitations appears.
While the precise behaviour depends on the details of the systems, a phenomenological trend was put forward showing a universal scaling $\Omega$/2$\Delta$ $\approx$ 0.64 \cite{Yu} which was later on tempered by considering carefully multi-gap systems systems and strong-coupling effects \cite{Inosov}. It was also stressed that the resonance arises from interplay between Fermi surface and superconducting gap topologies and is not specific to singlet or triplet superconductivity \cite{Morr,Yakiyama}. Concerning the local picture of the resonance mode, it has been proposed that it is a magnon mode originating from the proximity of a magnetic instability. In the normal state, the mode is overdamped and it is revealed in the superconducting state due to the suppression of the damping below 2$\Delta$ \cite{Morr2,Chubukov}. 
Finally a third case describes the situation where a well-defined low energy mode already exists in the normal phase, typically a dispersive crystal field level for $f$-electron systems, called crystal field exciton (unfortunately the same name exciton is used for the collective mode in the itinerant model). The interplay between this mode and the fermionic excitations can give rise to a satellite feedback peak at low energy in the superconducting phase \cite{Chang}. 

\begin{table*}
\caption{Heavy fermion compounds exhibiting  feedback effect of the superconductivity on the magnetic excitation spectrum at the wave-vector $\bf{k}$.\\ Relation between the superconducting transition temperature $T_{sc}$, the INS resonance energy $\Omega$, the damping rate of the resonance in the superconducting state $\Gamma_{sc}$, the superconducting gap 2$\Delta$ and the ratios $\Omega/k_{B}T_{sc}$, $\Omega$/2$\Delta$ and $\Gamma_{sc}$/$\Omega$. Data are taken from the INS references given in the main text. For the damping rate, when it is not available in the text of the references, a value inferred from the figures (half width at half maximum of the peak) is given and is preceded by $\sim$ in the Table. For the superconducting gap, the values for the maximum gap obtained by STM measurements are given and the references are indicated in the Table (except for UBe$_{13}$, where the value obtained by break-junction experiment is taken).}
\label{t1}
 \begin{adjustbox}{width=\textwidth}
\begin{tabular}{ccccccccccc}
\hline
&space group&$\bf{k}$& $T_{sc}$ (K)& $\Omega$ (meV) & $\Gamma_{sc}$ (meV) &2$\Delta$ (meV)&$\Omega/k_{B}T_{sc}$&$\Omega$/2$\Delta$ & $\Gamma_{sc}$/$\Omega$ \\
\hline
CeCoIn$_{5}$ &tetragonal & ($\approx$ 0.46, $\approx$ 0.46, 1/2)& 2.3 & 0.60 & $<$ 0.07  & 1.2 \cite{Allan} & 3.0 & 0.5 & $<$ 0.1\\
CeCu$_{2}$Si$_{2}$ &tetragonal & (0.215, 0.215, 1.458)& 0.6&  0.2 & 0.22  & 0.15\cite{Enayat} & 3.9 & 1.3 & 1.1\\
UPd$_{2}$Al$_{3}$$^{*}$ &hexagonal & (0, 0, 1/2)& 1.9& 0.36 &$\sim$ 0.15 & 0.47 \cite{Jourdan} & 2.2 & 0.8 & $\sim$ 0.4\\
UBe$_{13}$ &cubic & (1/2, 1/2, 0)& 0.85&  0.55  & $\sim$ 0.15  & 0.29\cite{Moreland} & 7.5 & 1.9  & $\sim$ 0.3\\
UTe$_{2}$ &orthorhombic & (0, 0.57, 0)& 1.6&   1.0 &0.68  & 0.5 \cite{Jiao} & 7.2 & 2.0 & 0.7\\
\hline
\end{tabular}
 \end{adjustbox}
\scriptsize\note{$^{*}$UPd$_{2}$Al$_{3}$ is magnetically ordered below  $T_{N}$=14.3 K with the propagation vector (0, 0, 1/2).}
\end{table*}

Table \ref{t1} summarizes the characteristic wave-vectors and energies obtained by INS in HF superconductors together with estimates of the superconducting gaps.
Already for the most extensively studied system CeCoIn$_{5}$ \cite{Stock, Raymond,Song}, the question about the resonance being a collective spin exciton \cite{Michal,Eremin} or a magnon mode\cite{Song,Chubukov} is not settled.
In this system, the resonance mode is sharp $\Gamma_{sc}$/$\Omega$ $<$ 0.1, intense and with $\Omega/k_{B}T_{sc}$ $\approx$ 3. The strong link between the magnetic resonance and the superconductivity was uniquely demonstrated by studies on doped CeCoIn$_{5}$. For Ce$_{1-x}$La$_{x}$CoIn$_{5}$ \cite{Panarin}, Ce$_{0.95}$Nd$_{0.05}$CoIn$_{5}$ \cite{Mazzone} and Ce$_{1-x}$Yb$_{x}$CoIn$_{5}$ \cite{SongYb} the decrease of $\Omega$  follows the one of $T_{sc}$ with essentially constant $\Omega/k_{B}T_{sc}$.
In the case of CeCu$_{2}$Si$_{2}$ ($T_{sc}$=0.6 K), the feedback of superconductivity on the excitation spectrum manifests through the opening of a gap while the broad signal above the gap is maintained \cite{Stockert}. It is found that $\Omega/k_{B}T_{sc}$ $\approx$ 3.9.
The model of interplay between fermionic degrees of freedom and crystal field exciton was specifically developed for UPd$_{2}$Al$_{3}$ in relation with the dual nature of $f$ electrons. A clear resonance occurs below the crystal field exciton \cite{Metoki,Bernhoeft,Sato} with $\Omega/k_{B}T_{sc}$ $\approx$ 2.2. For completeness, we mention the unique case of the HF superconductor PrOs$_{4}$Sb$_{12}$ where the suppression of damping reveals a clear crystal field exciton below $T_{sc}$ right at the energy 2.5$k_BT_c$ \cite{Kuwahara}. Despite the unconventional nature of the superconductivity, this looks like a classical crystal field effect.

In all these systems, the mode in the superconducting phase has a well-defined energy $\Omega$ in the range 2-4$k_BT_{sc}$.
In the less studied system UBe$_{13}$ ($T_{sc}$=0.85 K), an inelastic mode is already clearly evidenced in the normal state at about 0.55 meV and the feedback of superconductivity manifests by a  transfer of spectral weight from low to high energies while keeping the mode at around 0.55 meV and with a high value $\Omega/k_BT_{sc}$ $\approx$ 7.5.
Our findings of a redistribution of spectral weight in UTe$_{2}$ as well as a high ratio $\Omega/k_BT_{sc}$ have similarities with the behaviour of UBe$_{13}$.
The broad nature of the signal with $\Gamma_{sc}$/$\Omega$ $\approx$ 0.7 can be seen as a natural consequence of the high value of $\Omega/k_BT_{sc}$ when the mode lies above the superconducting gap.
This situation of a damped magnetic mode suggests a strong coupling scenario for the resonance\cite{Hlobil}.
The strong coupling for UTe$_{2}$ is also supported by the specific jump at $T_{sc}$, being higher than the BCS value \cite{Cairns}.
Contrarily to UBe$_{13}$, there is no identified well-defined mode in the normal phase of UTe$_{2}$ but a continuum of excitation described by a quasielastic response of relaxation rate $\Gamma_n$ \cite{Knafo}. 
A characteristic temperature $T^{*}$ $\approx$ 15 K separates a temperature independent regime from a temperature dependent regime for the staggered susceptibility at $\bf{k_1}$ and the associated relaxation rate with $k_{B}T^{*}$ $\approx$ $\Gamma_n$/2.
As pointed out in Ref.\citen{Knafo}, many thermodynamic and transport measurements exhibit an anomaly around $T^{*}$.
Interestingly, the new mode in the superconducting phase arises at $\Omega$ $\approx$ $k_{B}T^{*}$.  
This could also suggest a strong-coupling scenario in the frame of theoretical works showing that, in such a case, $\Gamma_{n}$ is the leading order term in the computation of $\Omega$, while it is 2$\Delta$ for weak-coupling \cite{Abanov,Manske}.
Last but not least, it should be stressed that compared to all other HF superconductors, UTe$_{2}$ realizes the only case where a true low-dimensional nature of the fluctuations is evidenced by INS.

To conclude, the feedback of the superconductivity on the magnetic excitation spectrum of UTe$_{2}$ manifests through a redistribution of spectral weight at the incommensurate wave-vector $\bf{k_1}$ forming a mode at 1 meV with a consequent broadening linked to a high ratio $\Omega/k_BT_{sc}$ $\approx$ 7.2. The wave-vector dependence along the $c$-axis show that the low dimensional behaviour characteristic of the normal phase with fluctuations mostly along the $a$-axis is essentially maintained in the superconducting phase.

\begin{acknowledgment}
This work was partly supported by the ANR grant FRESCO No. ANR-20-CE30-0020 and by Grants-in-Aid for Scientific Research (C) (No. 19K03756), (B) (No. 20H01864) and (S) (No. 21H04987) 
from the Japan Society for the Promotion of Science. 
The neutron scattering data collected at ILL for the present work are available at https://doi.ill.fr/10.5291/ILL-DATA.INTER-546.
We thank H. Suderow for scientific discussion.
\end{acknowledgment}

\end{document}